\documentclass{IEEEtran}

\usepackage[table]{xcolor}
\usepackage{cite}
\usepackage{graphicx}
\usepackage{lipsum}
\usepackage{stfloats}
\usepackage{amssymb}
\usepackage{pifont}
\usepackage{bbding}
\usepackage{booktabs}
\usepackage{ctable}
\usepackage{threeparttable}
\usepackage{footmisc}
\usepackage{framed} 
\usepackage{array}
\usepackage{multirow}
\usepackage{longtable}
\usepackage{rotating}
\usepackage{fancyhdr}
\usepackage{lastpage}
\usepackage{layout}
\usepackage{wrapfig}
\usepackage{xcolor}
\usepackage{tabularx}
\usepackage{tabu}
\usepackage{enumitem}
\usepackage{url}
\graphicspath{{Figures/}}
\DeclareGraphicsExtensions{.png, .eps}

\begin{document}
	\title{Generative AI for Semantic Communication: Architecture, Challenges, and Outlook}
	\author{
	Le Xia, Yao Sun, Chengsi Liang, Lei Zhang, Muhammad Ali Imran, and Dusit Niyato
	\thanks{
	
	Le Xia, Yao Sun (\textit{corresponding author}), Chengsi Liang, Lei Zhang, and Muhammad Ali Imran are with University of Glasgow, United Kingdom;
	
	Dusit Niyato is with Nanyang Technology University, Singapore.
	
	}}	
	\maketitle
	\begin{abstract}
		Semantic communication (SemCom) is expected to be a core paradigm in future communication networks, yielding significant benefits in terms of spectrum resource saving and information interaction efficiency.
		However, the existing SemCom structure is limited by the lack of context-reasoning ability and background knowledge provisioning, which, therefore, motivates us to seek the potential of incorporating generative artificial intelligence (GAI) technologies with SemCom.
		Recognizing GAI's powerful capability in automating and creating valuable, diverse, and personalized multimodal content, this article first highlights the principal characteristics of the combination of GAI and SemCom along with their pertinent benefits and challenges.
		To tackle these challenges, we further propose a novel GAI-integrated SemCom network (GAI-SCN) framework in a cloud-edge-mobile design.
		Specifically, by employing global and local GAI models, our GAI-SCN enables multimodal semantic content provisioning, semantic-level joint-source-channel coding, and AIGC acquisition to maximize the efficiency and reliability of semantic reasoning and resource utilization.
		Afterward, we present a detailed implementation workflow of GAI-SCN, followed by corresponding initial simulations for performance evaluation in comparison with two benchmarks.
		Finally, we discuss several open issues and offer feasible solutions to unlock the full potential of GAI-SCN.
		
	\end{abstract}
	
	\section{Introduction}
	Recently, semantic communication (SemCom) is popularized as an emerging paradigm that promises to significantly alleviate the scarcity of communication resources in future wireless networks~\cite{huang2021deep}.
	This is mainly benefited from prosper advancement in deep learning (DL) technologies that can drive semantic encoding and decoding models to achieve efficient and high-quality semantic refinement on desired meaning with low spectrum consumption.
	Moreover, through equipping both ends of the transceiver with equivalent background knowledge, the implicit meaning in conveyed content can be recovered with ultra-low semantic errors even under harsh channel conditions.
	However, realizing such superiorities obviously poses a huge demand on data acquisition for constructing background knowledge and pre-training DL-driven semantic models.
	Meanwhile, considering that the achievable semantic performance is essentially confined by the quality of pre-training data used, existing SemCom systems still lack sufficient context reasoning capabilities, i.e., accurate semantic calibration and recovery in transmitting multiple complex and coherent contextual fragments.
	
	Fortunately, state-of-the-art (SOTA) generative AI (GAI) models, have lately emerged as killer applications in many verticals, promising to bring considerable productivity, innovation, and economic value to the real-world services as diverse as image synthesis, text generation, and drug discovery~\cite{xu2024unleashing}.
	To be concrete, GAI leverages powerful DL algorithms, such as Transformer and diffusion to automate photorealistic and multimodal AI-generated content (AIGC) in response to user-provided prompts, while its fidelity and accuracy are contingent on adequate pre-training on billions of parameters in large language (e.g., ChatGPT) or image (e.g., Dall-E) models.
	Most importantly, GAI has immense abilities of context-reasoning and cross-modal content synthesis to generate high-quality and basically correct responses by successfully mimicking human's thinking and speaking patterns, enabling users feel that they are interacting with a real human-being rather than a dull machine~\cite{erdemir2023generative,kingma2019introduction,wang2022diffusiondb,rezende2015variational}.
	Such a milestone innovation stimulates us to investigate the potential of applying AIGC into wireless SemCom, which hypothesizes to yield the below benefits.
	
	\textbf{Better SemCom Training Efficiency:}
	Undoubtedly, the growing prosperity of SemCom is inseparable from colossal data resources for semantic model pre-training and background knowledge preparation targeting effective semantic interpretation~\cite{xia2023wiservr}.
	To enable better SemCom training efficiency, GAI models are capable of producing vast multimodal content with a certain degree of authenticity and thus should be valuable materials to semantic training and background knowledge.
		
	\textbf{Enhanced Semantic Context-Reasoning:}
	The historical AIGC can be stored online and retrieved easily to provision semantic coding models a better understanding for the context information, thereby offering significant context reasoning and semantic generalization.
	Furthermore, the creativity of GAI models can offer high-quality and precise content automatically for semantic interpretation in SemCom, thanks to the Prompt Engineer mechanism~\cite{liu2022design}.
	This ensures high semantic fidelity even if coding errors occur during joint-source-channel coding (JSCC) or physical signal transmission.
		
	\textbf{Higher Spectrum Utilization:}
		Notice that most of AIGC can be produced via inputting only a few prompts, and the responses can be highly precise and specific if the prompts are well-crafted to align with a task-oriented communication.
        Hence, by utilizing well-trained GAI models, only several prompts need to be sent, instead of transmitting the whole source information, in each SemCom process to significantly deduct the required bandwidth resources while retaining the original meaning.

        Despite many ascendancies offered by the combination of GAI and SemCom, it still encounters several inevitable and thorny challenges in practical implementation.
        Among them, the paramount issue is how to deal with such considerable computing and storage resources required by these large GAI models.
        For instance, OpenAI's product ChatGPT-3 comprises approximately 175 billion parameters in total~\cite{xu2024unleashing}, and hence it undoubtedly needs to consume colossal computing resources to operate the system.
        Another problem worth pointing out here lies in the reliability and latency aspects.
        AIGC is autonomously created that may lead to uncertainty to some extent, while introducing extra delay for data processing as well as data dissemination.
		To the best of our knowledge, only a few studies have explored the potential of incorporating GAI with SemCom.
		For example, the authors in~\cite{erdemir2023generative} and~\cite{thomas2023neuro} employed different deep GAI networks to enhance the perceptual quality and semantic reliability for SemCom.
		Similarly,~\cite{guo2023semantic} and~\cite{barbarossa2023semantic} respectively sought the possibility of using GAI to quantify the semantic importance and adapt the end-to-end transmission rate.
		However, all of these works consider it from a transceiver design perspective in device-to-device scenarios, while relevant research for the system framework design in general cellular networks is still lacking.
      	To this end, this article focuses on unleashing the full potential of GAI-integrated SemCom network (GAI-SCN) across the cloud-edge-mobile layers, and the main contributions are summarized in a nutshell as follows:
	\begin{itemize}
		\item We first present the basic concept of SemCom and four major types of GAI technologies, and then provide a comprehensive comparison among traditional communication, SemCom, and GAI-integrated SemCom. Next, we present the potential junctions between SemCom and GAI.
		\item We propose a novel GAI-SCN framework that integrates global and local GAI with semantic coding models in a collaborative cloud-edge-mobile design. Afterward, we showcase its viable implementation workflow consisting of three successive stages: Initial Network Preparation Stage, GAI-integrated SemCom Service Provisioning Stage, and Model Synchronization and Update Stage.
		Moreover, numerical results validate that the proposed framework can save a significant number of transmitted bits while maintaining high-precision semantic delivery compared with two benchmarks.
		\item Finally, several open issues with prospects of GAI-SCN are outlined, including device hardware limitations, inactive information sharing of users, and potential data tampering and privacy leakage.
	\end{itemize}
	
	\section{When SemCom Meets GAI}
\begin{figure}[t]
		\centering
		\includegraphics[width=0.495\textwidth]{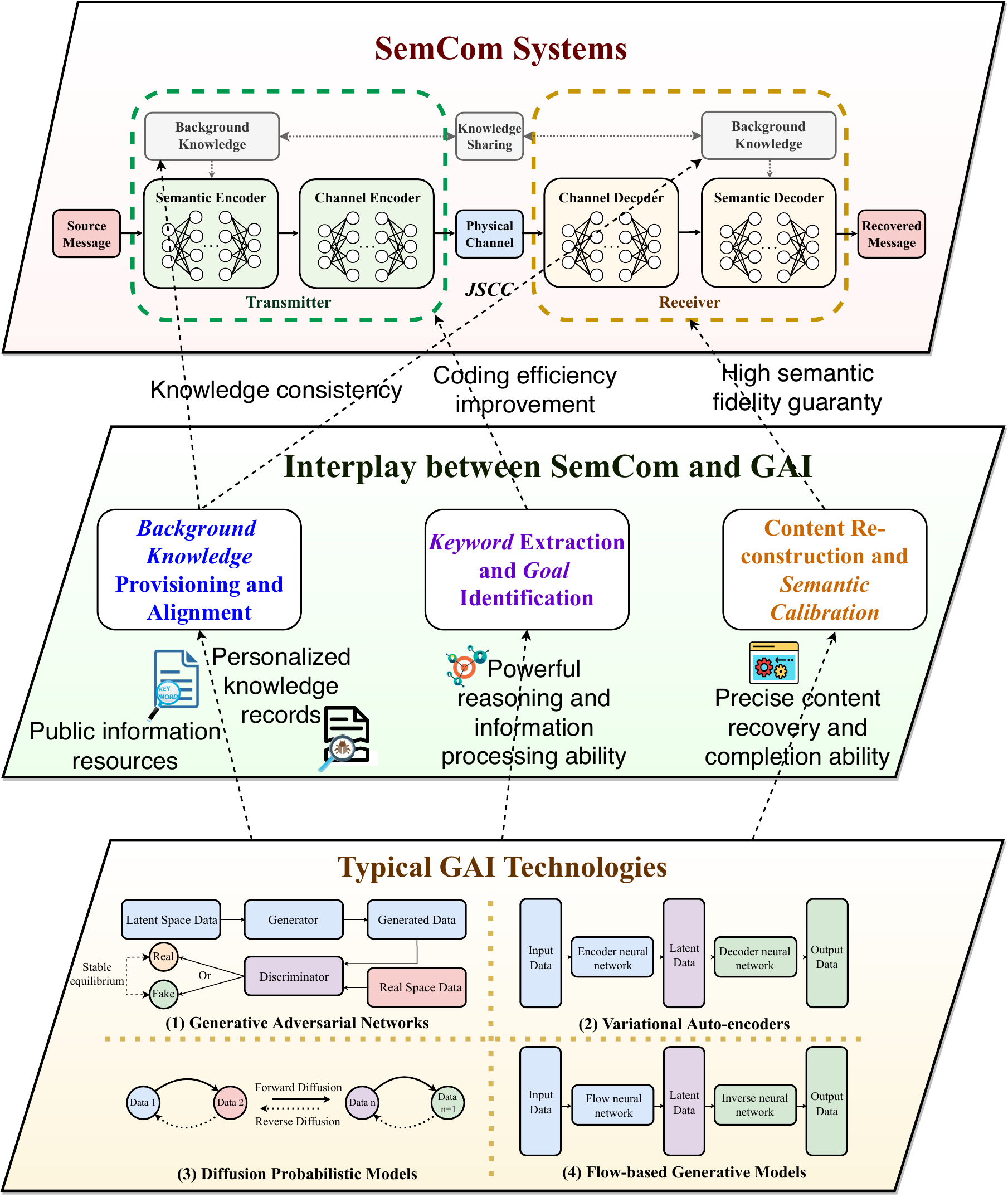} 
		\caption{Overview of SemCom systems and four types of typical GAI technologies along with three aspects of interplay between SemCom and GAI.}
		\label{Scenario}
    \end{figure}

        In this section, we first introduce typical technologies of SemCom and GAI, as shown in Fig.~\ref{Scenario}, followed by several junctions between them identified and discussed in detail.
        
	\subsection{SemCom Systems}
        Compared with traditional communication of guaranteeing the precise reception of transmitted bits, the accurate delivery of semantics implied in desired messages becomes the cornerstone of SemCom~\cite{barbarossa2023semantic}. Taking an end-to-end SemCom system as the example, a transmitter first leverages background knowledge relevant to source messages to filter out irrelevant content and extract core features that only require fewer bits for transmission, the process of which is called semantic encoding. Once the receiver has the required knowledge, its local semantic interpreters are capable of accurately restoring the original meanings from the received bits, even with intolerable bit errors in data dissemination. This process is called semantic decoding. Consequently, efficient exchanges for the desired information with ultra-low semantic ambiguity can be achieved in SemCom under equivalent background knowledge, while significantly alleviating the resource scarcity problem.
        
        \subsection{Typical GAI Technologies}
        With proper pre-training and fine-tuning alongside extensive datasets, GAI excels in learning background knowledge and content structures from input training data, thereby generating outputs that closely resemble real-world samples~\cite{xu2024unleashing}.
         In what follows, we briefly introduce four basic GAI technologies: generative adversarial networks (GANs), variational auto-encoders (VAEs), diffusion probabilistic models (DPMs), and flow-based generative models (FGMs).
         Note that other technologies like Transformer (related to ChatGPT) or the variants of these four (related to Dall-E) are equally essential as the indispensable components in the most of existing mainstream GAI models.
        
        \textbf{GANs:} 
		The GAN consists of two distinct neural networks: a generator and a discriminator.
        In the form of constant contestation, the goal of generator is to confuse the discriminator, while the discriminator should distinguish the samples generated by the generator from the real samples, until reaching a stable equilibrium~\cite{erdemir2023generative}.
        Although the GAN enables GAI-SCNs to output a certain degree of accurate content, it is still not satisfactory enough in terms of semantic quality, generation diversity, and multimodal distribution problem learning.
        
	\textbf{VAEs:}
        The VAE is a likelihood-based generative auto-encoder model, normally comprising of a multi-layer encoder and a symmetric decoder.
        By taking random training samples with a specific distribution as input, the encoder can regularize its coding distribution to ensure good properties of latent space, while the decoder maps from the latent space to the input space so as to produce new data points~\cite{kingma2019introduction}.
        When it comes to the GAI-SCNs, efficient multi-task SemCom is envisioned to be realized with the support of VAEs.
	
	\textbf{DPMs:}
        The DPM is a class of latent variable models, and its generation principle mainly includes two processes of forward diffusion and reverse diffusion.
        In the forward process, the input content is polluted in steps by introduced Gaussian noise.
        In comparison of GANs and VAEs, DPMs render remarkable superiority in semantic recovery tasks (e.g., image denoising, inpainting, and super-resolution), producing high-quality content and have better resistance to the risk of noise and interference~\cite{wang2022diffusiondb}.
        
        \textbf{FGMs:}
        Differing from the previous models, FGMs are exact log-likelihood models with tractable sampling and latent variable inference, which applies a bunch of reversible transformations to samples from the prior so that log-likelihoods of observations can be computed~\cite{rezende2015variational}.
        Besides, it leverages the change-of-variable law of probabilities to transform a simple distribution into a complex one, which greatly facilitates the semantic generation accuracy and communication efficiency.
	
	\subsection{Interplay between SemCom and GAI}
	Based on the above introduction of SemCom systems and GAI technologies, herein we list three principal conjunctions between them with the corresponding elaboration.

    \textbf{Background Knowledge Provisioning and Alignment:} 
    	Technically, GAI can be exploited as valuable data assets for users' background knowledge provisioning, which is roughly divided into two categories of global knowledge and personalized knowledge.
        For starters, global knowledge represents the common information publicly available to society (e.g., the content recorded in books, articles, videos, and other online sources), while personalized knowledge indicates users' personal information (e.g., language habits and communication style preferences).
        Thanks to sufficient pre-training, existing large GAI models like ChatGPT can easily and quickly retrieve global knowledge online to be stored as the common background knowledge of users.
        More importantly, such AI-generated global knowledge guarantees information consistency between any pair of communication parties, which ensures knowledge equivalence in SemCom.
        As for the personalized knowledge, it can be stipulated that GAI models store private conversations with users in the preparation stage of SemCom, by which their preferences can be analyzed in the background so as to offer personalized and customized AIGC according to local environments.
            
	\textbf{Keyword Extraction and Goal Identification:} 
		Recap that the core of SemCom is meaning delivery, for this purpose, GAI models are capable of extracting some keywords from the input long content, and the corresponding communication goal can be identified in a small amount of text (or the bounding box of objects in image) through excellent context-understanding ability.
        {Note that the extracted semantic granularity with respect to keywords and goals can be flexibly adjusted according to the importance of semantic information combined with each user's personal preference, especially in the task-oriented semantic transmission scenario.
        Accordingly, the global GAI model (e.g., a stable diffusion model) is trained specifically to restore the original content from received keywords and goals given background knowledge.
        Moreover, fewer communication resources (including wireless bandwidth and energy) are demanded, while the pressure from stringent latency requirements is relieved.
        
	\textbf{Content Reconstruction and Semantic Calibration:} 
        Cooperating with the keyword extraction function at the transmitter side, the GAI models deployed either in the core network or on the receiver side can realize content reconstruction according to different SemCom goals.
        Besides, a certain degree of semantic ambiguity may inevitably arise in the process of signal transmission and semantic interpretation, such as wording or sentence structure errors in the delivered text and blurred or partially missing tiles in the delivered images.
        To this end, the initially recovered content after semantic decoding can be input into some GAI models (like GPT-Neo) for fundamental and comprehensive semantic calibration to further improve the accuracy and reliability of SemCom.
        
    \textit{\textbf{Lesson Learned:}} The GAI is promising to be integrated with SemCom especially for task-oriented (i.e., meaning delivery-driven) and high-capacity modern entertainment communication services, such as mobile virtual reality delivery and Metaverse.
    Notably, although producing AIGC can consume some extra local processing time, the transmission delay is greatly reduced by SemCom in parallel, promising to significantly relieve the latency-cost pressure.
    Furthermore, we specially compare the system characteristics of GAI-integrated SemCom with traditional communication and SemCom alongside their respective benefits and limitations, as sketched in Fig.~\ref{Three}.
    Among them, traditional communication is built on bit-based source-and-channel coding following pre-defined and stringent codebooks, while SemCom integrates AI-based semantic encoder and decoder with equivalent background knowledge for accurate semantics-aware JSCC.
    On this basis, GAI-integrated SemCom takes full advantage of GAI technologies to significantly enhance semantic delivery efficiency and ease resource pressure.  
        
    \begin{figure}[t]
		\centering
		\includegraphics[width=0.495\textwidth]{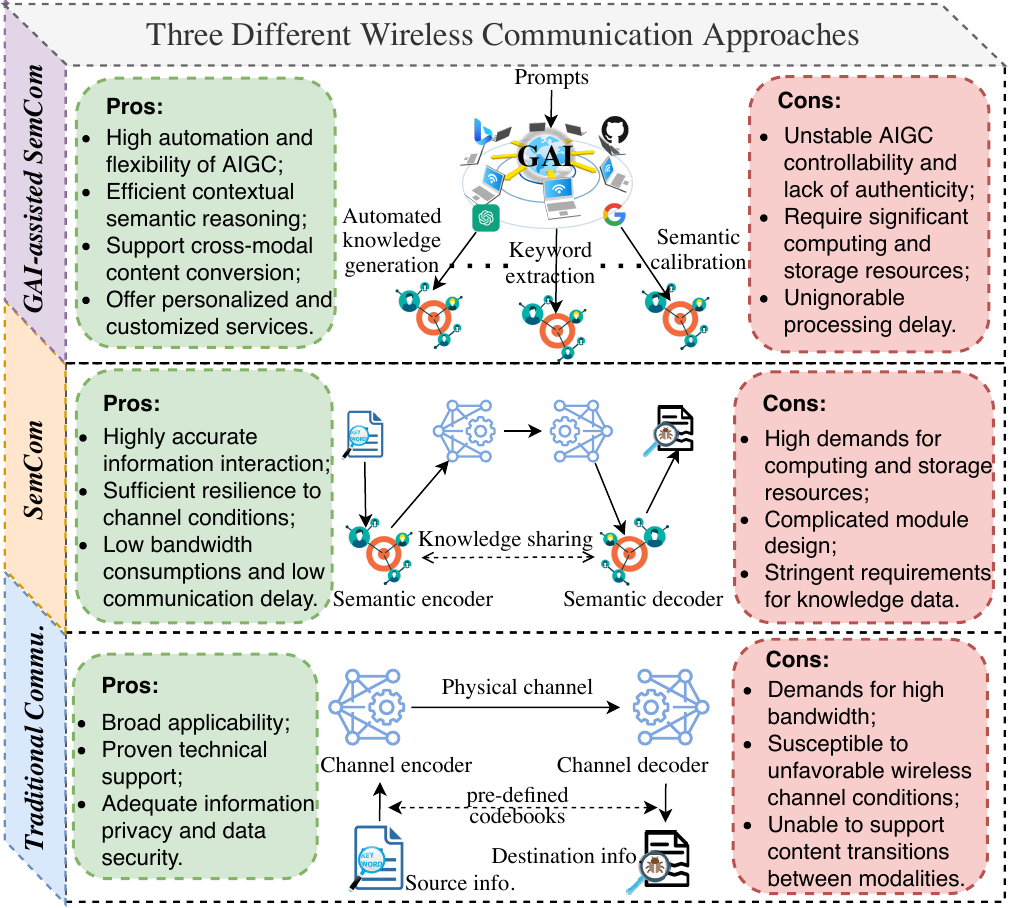} 
		\caption{Comparisons among three different approaches of GAI-SemCom, SemCom, and traditional communication in terms of their pros and cons.}
		\label{Three}
    \end{figure}
        
       
	\section{GAI-Integrated SemCom Network Framework}
	
	\subsection{Hierarchical Structure of GAI-SCN}
     	Consider a SemCom-enabled cellular network scenario as demonstrated in Fig.~\ref{GAI-SCN}, where there are multiple terminal devices (TDs) of senders and receivers within the coverage of base stations (BSs).
   	    Among them, multimodal SemCom services (e.g., text, image, and video) with specific communication goals consecutively arrive at each TD, and each BS acts as the service controller to efficiently schedule and coordinate the goal-oriented SemCom service provisioning.
     	Additionally, a large GAI model (e.g., GPT-4 or Dall-E) is deployed in the cloud to complete computationally intensive tasks, while a small one (e.g., GPT-Neo~\cite{kashyap2022gpt}) is embedded in the TD for coping with local lightweight service demands, such as customized content extraction and text generation services.
        
	\begin{figure*}[ht]
		\centering
		\includegraphics[width=0.985\textwidth]{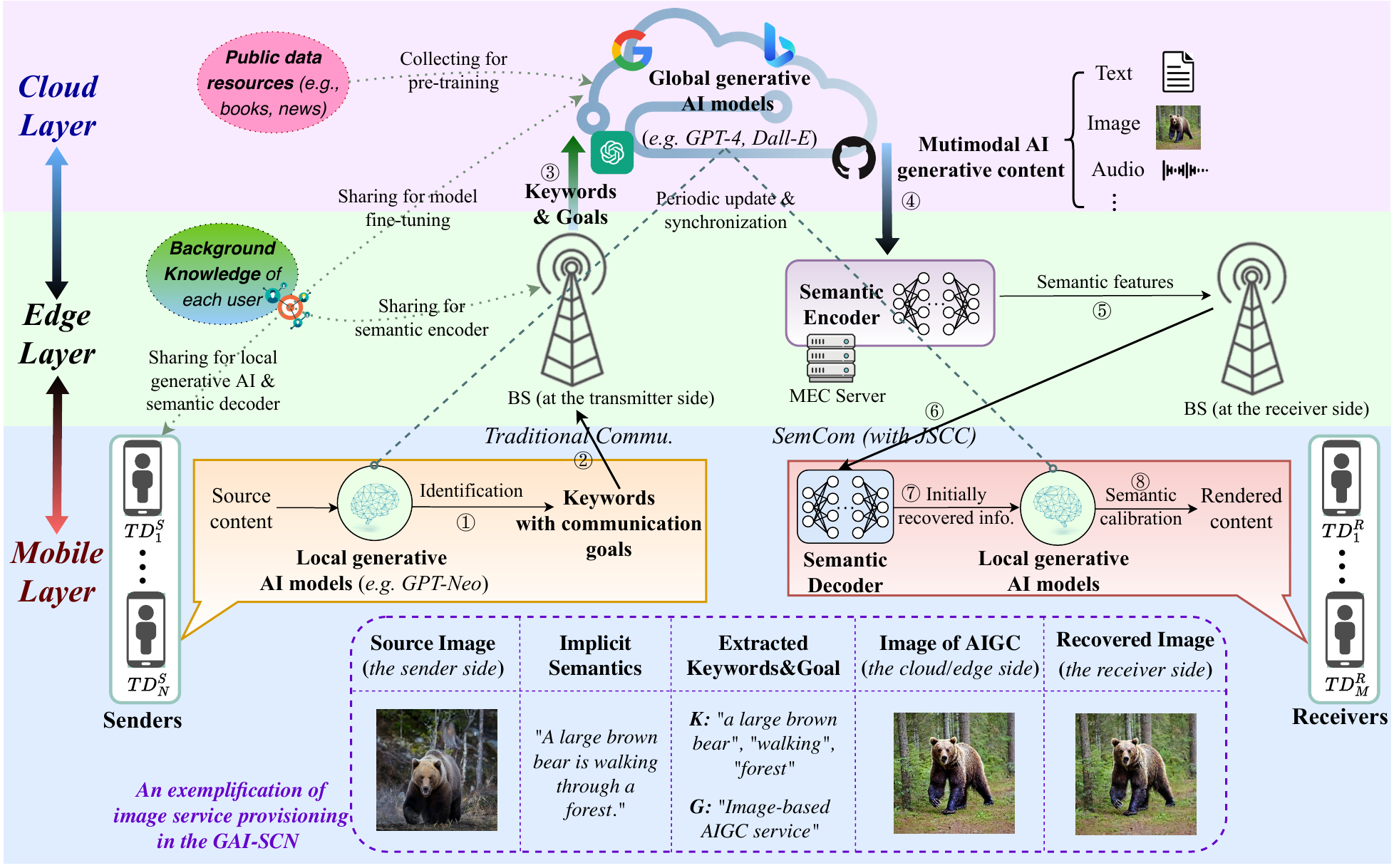} 
		\caption{Illustration of the proposed GAI-SCN framework in a collaborative cloud-edge-mobile design, where an exemplification of image service provisioning is presented.}
		\label{GAI-SCN}
    \end{figure*}
	
	\textbf{Service Provisioning in the Mobile Layer:}
	Each TD is first equipped with a light-weight GAI model to realize context-aware keyword extraction and goal identification, and compared to data compression methods, this takes full advantage of the context reasoning ability of GAI to guarantee more efficient and intelligent semantic transmission.
	Through pre-training on large datasets (like Wikipedia and Common Crawl) and fine-tuning on user data (to be more personalized and customized), GPT-Neo is capable to be directly installed into TDs to precisely extract keywords of source information for users and identify the communication goal in only several words, by which fewer bits and smaller latency are consumed on the transmitter side for data transmission.
	As for the TDs on the receiver side, semantic decoders are deployed to recover the delivered meaning from obtained bits, after which the GPT-Neo can be further used for semantic calibration or language translation, etc.
	Note that the GPT-Neo can be flexibly replaced by other SOTA DL-based semantic extraction models, as long as they can efficiently accomplish the task of extracting keywords and goals accurately.
		
	\textbf{JSCC Process in the Edge Layer:}
	By enabling sufficient computing and storage ability in distributed edge servers, semantic encoders are considered to be deployed in the edge layer of GAI-SCN.
	On one hand, we exploit the JSCC method as it is able to greatly improve resilience and robustness of SemCom especially against the extreme channel conditions, where the essential processes like pre-training, fine-tuning, and reasoning can be completely handled by edge computing servers.
	On the other hand, considering the large-capacity AIGC transmission scenario that requires excessive wireless network resources, SemCom can ensure not only the resource savings but also the accurate semantic conveying and recovery for the application-layer counterpart, where related semantic coding tasks can be also offloaded to edge servers.
	
	\textbf{AIGC Acquisition in the Cloud Layer:}
	Based on the above mobile and edge layer designs, the generation of AIGC becomes an indispensable process for smooth SemCom.
	In the proposed GAI-SCN, a centralized infrastructure, i.e., the remote cloud server, can support and run large GAI models like Google Bard or Microsoft Bing Chat.
	All preparation processes related to the AI model itself, such as pre-training and fine-tuning, are accomplished in the cloud.
	Likewise, the collection, analysis, and reasoning for multi-users' personal information and preferences can be realized in real time by leveraging the massive computing and storage resources of the cloud servers.
	Moreover, according to the specific keywords and goals uploaded from sender TDs, these large GAI models can quickly and correctly create the response content of desired modes.
	The rationale behind this is due to the historical conversation record, i.e., referred to as context, between users and GAI servers, which provides reference information for models' understanding and inference.
	Therefore, the original meaning implied in the delivered keywords can be recovered to intact text or image content.
	
	\textbf{\textit{Lesson Learned:}} By adopting the above collaborative cross-layer GAI-SCN framework, both link-level wireless resource utilization and application-layer performance (e.g., satisfaction of user semantic provisioning) can be maximized.
	Particularly, JSCC-based SemCom is a very appropriate solution to ensure not only resource savings but also accurate key semantics conveying and recovery for the application-layer counterpart.
	Note that our proposed GAI-SCN does not assume the common phenomenon of signal transmission inconsistency in SemCom, but instead, its whole design is based on the consideration of how to accurately transmit and recover the core semantics between the source and destination.
	Besides, although the usage cost of GAI-SCN is higher than traditional communication or general SemCom, the savings in communication resources (e.g., spectrums and communication delay) and the promotion in communication quality (e.g., semantic transmission efficiency and resilience to signal distortion) should be considerable as well.
	Most importantly, the global GAI is placed in the cloud that is generally computing-resource-unlimited, while semantic coding models are preset as maturely trained in advance that means the related training costs are not counted into the budget.
	
	\subsection{Implementation Workflow}
	By revisiting the key rationales of GAI-SCN, we now showcase its workflow below step-by-step to guide network designers to make proper changes on relevant protocols, as shown in Fig.~\ref{Implementation}.
    \begin{figure*}[ht]
		\centering
		\includegraphics[width=0.95\textwidth]{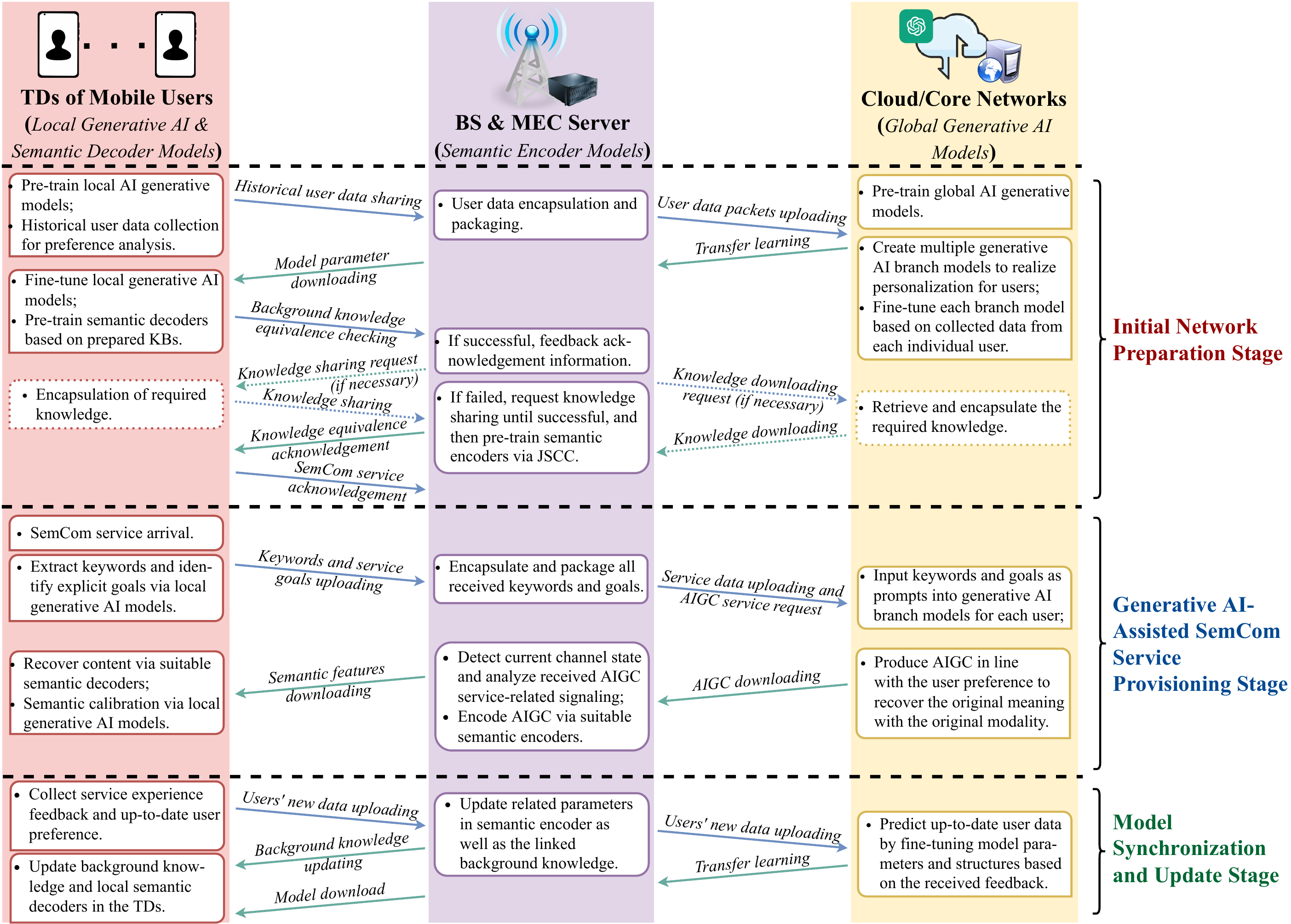} 
		\caption{A schematic diagram of implementing a complete round of semantic service provisioning in the GAI-SCN, including three successive stages of Initial Network Preparation, Generative AI-integrated SemCom Service Provisioning, and Model Synchronization and Update.}
		\label{Implementation}
    \end{figure*}
        
	\textbf{Initial Network Preparation Stage:} 
	In this stage, the construction, pre-training and fine-tuning of GAI models are implemented collaboratively across the cloud and mobile layers.
	Note that all AI models are well pre-trained under society data public to users in the first place.
	Subsequently, considering the preference discrepancy between different users, personal data are collected by local GAI and then shared with global GAI in the cloud.
	Next, multiple parallel branch models corresponding to multiple users are created and fine-tuned based on personal data, where the branch model refers to a design pattern about multiple paths or subnetworks to make different predictions.
	Afterward, all parameters trained in each branch model are downloaded into the local light-weight GAI model related to each user, so as to avoid the drawback of limited computing ability in mobile TDs.
	Apart from the above, the construction and joint pre-training of semantic encoders and decoders are also completed in advance given specific channel state information.
	Keeping in mind the requirement of knowledge equivalence in SemCom, knowledge sharing between encoders and decoders is required if the condition is triggered~\cite{xia2023wiservr}.
	
	\textbf{GAI-Integrated SemCom Service Provisioning Stage:}
	Once all communication parties in the GAI-SCN are equipped with mature-trained AI models, the SemCom transmission services begin.
	The local GAI model first extracts keywords and identifies the explicit goal (referring to the exemplification in Fig.~\ref{GAI-SCN}) for each user.
	Hereafter, such keywords and goals are uploaded via traditional communications to global GAI models in the cloud layer as input prompts, so that the corresponding branch AI model can create AIGC in line with the user preference to recover the original meaning with the original modality.
	
	When it comes to the downlink side, each edge server detects the current wireless channel state and analyzes the received AIGC service-related signaling to select an appropriate pre-trained semantic encoder.
	As such, the entire AIGC is smoothly encoded into semantic features to be transmitted to the corresponding TD.
	Since semantic errors may still occur in the above JSCC process due to potential signal impairment and the limited computing power of TD, under the assistance of personalized knowledge and user preference, its local generative-AI model is utilized for further semantic calibration, such as error correction for text content, color rendering for image content, and frame completion for video content, to enhance the resilience and robustness against semantic ambiguity for SemCom.
	
	\textbf{Model Synchronization and Update Stage:}
	At the end of each GAI-integrated SemCom process, the service experience feedback (i.e., the user satisfaction regarding service performance, availability, and accessibility, etc.) is collected from each user and cached in the associated edge server.
	The feedback data as well as real-time user transmission data are synced periodically from the edge layer to the GAI models deployed in the other two layers.
	In this way, GAI analyzes and predicts users' up-to-date preferences by fine-tuning model parameters and structures.
	Meanwhile, the data are also fed back to semantic models to update the relative parameters as well as the linked background knowledge.
	
	
	\section{Case Study: Image Transmission Service Provisioning in GAI-SCN}
	In this section, numerical results for a case study of image transmission are presented to evaluate the initial performance of the proposed GAI-SCN framework.
	For the simulation settings, an image-captioning model by combining ViT model with GPT-2 model~\cite{shen2023hugginggpt} is exploited as the local GAI to realize the image-to-text transformation as well as the keyword extraction and goal identification.
	Besides, we employ the latest text-to-image model called Stable Diffusion 2.1~\cite{wang2022diffusiondb} as the global GAI to create the AI-generated images from received prompts.
	As for the SemCom part, the main setups are proceeding as in the work~\cite{xia2023wiservr}, where an advanced deep convolutional network named Observation-Centric Sort and a Transformer-powered semantic decoder are leveraged for semantic segmentation and recovery, respectively. 
	Meanwhile, all semantic models are trained based on the additive white Gaussian noise channel with a signal-to-noise ratio (SNR) of $0$ dB to transmit $327$ images with different contents for testing.
	Finally, the Adam optimizer is adopted to train the neural networks in GAI-SCN with an initial learning rate of $5\times 10^{-4}$ based on the given image dataset.
	In parallel, for comparison purposes, we utilize two benchmarks: 1) A GAI-integrated traditional communication scheme, where the AIGC is encoded into bits based on the variable length source coding and LDPC channel coding~\cite{cai2006efficient} for precise image delivery; 2) A typical SemCom scheme~\cite{huang2021deep,erdemir2023generative}, which transmits the original images via only DL-based semantic coding and JSCC without any involvement of GAI.
	
	\begin{figure}[t]
		\centering
		\includegraphics[width=0.48\textwidth]{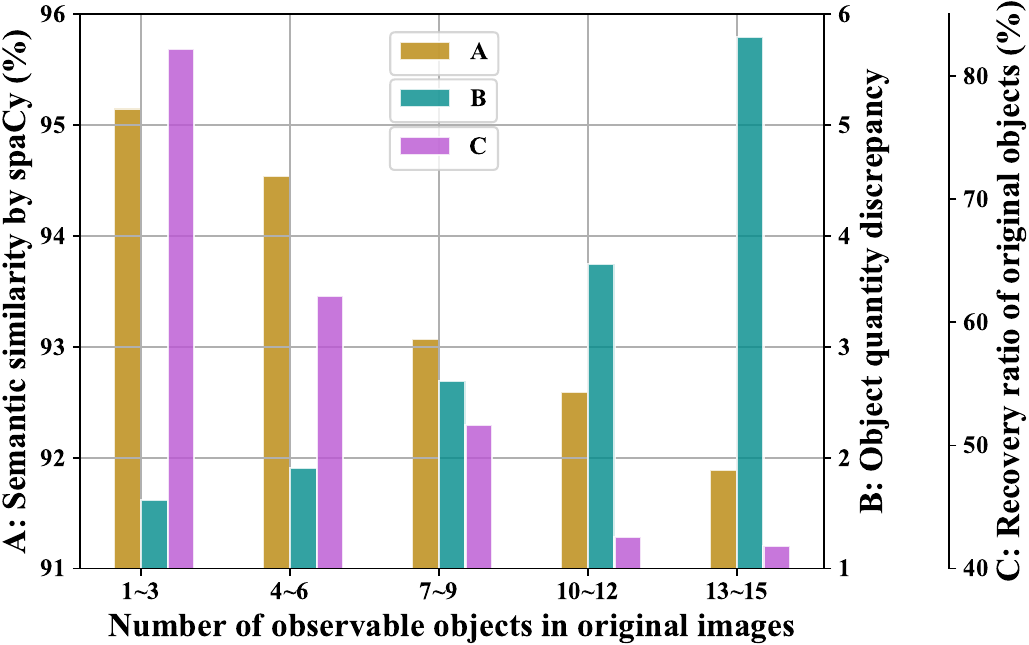} 
		\caption{Comparisons between original and recovered images by the proposed GAI-SCN framework in terms of three metrics: A) Semantic similarity by spaCy; B) Object quantity discrepancy; C) Recovery ratio of original objects.}
		\label{Figure4}
    \end{figure}
    
    Figure~\ref{Figure4} first demonstrates the performance of GAI-SCN by comparing the recovered images with the original ones under different numbers of observable objects that can be detected and segmented (such as ``bear'', ``tree'', and ``ground'' as shown in the exemplification image in Fig.~\ref{GAI-SCN}).
    Note that the appearances of the transmitted images may be not completely consistent with the received ones, however, the semantics implicit in the content should be our sole focus in measuring the system performance.
   	Herein, we first test the semantic similarity performance measured by spaCy~\cite{vasiliev2020natural}, where generally, the higher the spaCy score, the more accurate the recovered semantics.
   	Due to the cross-layer design of our proposed GAI-SCN framework, semantic similarity is actually an application-layer indicator to measure the degree of user semantic provisioning, which should be distinguished from lower-level technical indicators such as bit/symbol error ratios.
   	Particularly, the error occurs at the bit/symbol level does not necessarily imply an error at the semantic level, while there could be errors at the semantic level even if there is no error at the bit level. The former is owing to the powerful semantic reasoning ability of semantic models and the assistance of perfectly matching background knowledge, and the latter is because of the potential background knowledge discrepancy between the sender and the receiver.
   	Back to Fig.~\ref{Figure4}, it is observed that the increasing number of objects in original images results in a decreasing semantic similarity, which is because that the higher complexity of images makes GAI more difficult to extract the keywords correctly as well as the image recovery.
   	This phenomenon is also consistent with the recovery ratio performance of original objects, which metric shows the proportion of objects accurately recovered in the AIGC, and as the complexity of images rises, the average recovery ratio drops from $82.2\%$ to $41.8\%$ steadily.
   	Moreover, we can see a lower object quantity discrepancy with fewer number of objects contained in the transmitted images.
   	All of these trends above represent that semantic ambiguity is more likely to occur in regenerating more complex images due to confusing key object identification and a certain degree of semantic interference.
	
	\begin{table}[!t]
		\centering
		\caption{Amount of Bits Required and the PSNR Performance in Transmitting $300$ Images on the Downlink ($1024*1024$ Pixels)}
		\label{Time}
		\setlength{\tabcolsep}{3pt}
		\renewcommand\arraystretch{1.4}
		\begin{tabular}{|m{3cm}<{\centering}|m{2.5 cm}<{\centering}|m{2.5 cm}<{\centering}|}\hline
			\textbf{Different image transmission schemes} & \textbf{Number of required bits for downlink} & \textbf{PSNR}\\ \hline
			GAI-integrated traditional communication~\cite{cai2006efficient} & $1.28\times 10^{5}$ & $28.05$\\ \hline
			SemCom~\cite{huang2021deep} & $5.99\times 10^{4}$ & $28.25$\\ \hline
			GAI-SCN & $3.03\times 10^{4}$ & $28.64$\\ \hline
		\end{tabular}
	\end{table}
	
	Besides the above AI-generated quality measurement, the proposed GAI-SCN is further tested from the wireless transmission perspective.
	Table~\ref{Time} shows the number of bits required and the image transmission performance achieved to downlink $300$ images with the equal size of $1024*1024$ pixels in the same $0$ dB SNR Gaussian noise channel.
	For tractability, the peak signal-to-noise ratio (PSNR) is employed here to compare the images before and after performing different transmission schemes, noting that other image quality evaluation metrics like structural similarity index can also be adopted.
	Moreover, since we concentrate only upon the discrepancy in transmitted content (AIGC vs. original) and communication technology used (SemCom vs. traditional), the same source and channel coding rules are assumed for all schemes to encode the images from the pixel/semantic level to the bit level.
	It can be found that the proposed GAI-SCN only requires $3.03\times 10^{4}$ bits, which reduces $2.96\times 10^{4}$ bits compared with the SemCom scheme and $9.77\times 10^{4}$ bits with the GAI-integrated conventional scheme.
	Furthermore, the PSNR score obtained by our GAI-SCN maintains a very high level of $28.64$, which is even slightly better than the other two approaches.
	This can be explained by the fact that the typical SemCom scheme starts the image delivery process from the uplink direction, which increases the risk of image sharpness loss, while the GAI-integrated traditional communication scheme is less resilient to harsh channel conditions compared with GAI-SCN.
	In summary, the above results demonstrate that our GAI-SCN can further save bandwidth resources while guaranteeing very high-quality SemCom service provisioning with accurate semantic delivery.
	
    \section{Open Research Issues and Outlooks}
    In this section, we list several thorny issues and outlooks that can be highlighted as future research directions in the GAI-SCN.
    
    \textbf{Limited Device Resources for Supporting AI Modules:}
    In the GAI-SCN, sophisticated AI-enabled computing modules (including local GAI models and SemCom coding models) need distributed implementation at each TD, imposing a heavy burden on its inherently limited device resources (like storage, memory, computational units, and battery power).
    To make it practically implementable, advanced model compression and acceleration technologies, such as knowledge distillation, parameter pruning and quantization, are promising to efficiently drop the complexity and size of AI networks with an affordable cost of performance degradation.
    
   \textbf{Randomness of Content Rendered in GAI-SCN:}
   Notice that the appearance of AIGC output from the cloud GAI may vary even given the same keywords and goals.
   Besides, the representation of semantics recovered via the semantic decoder also have uncertainty to some extent, due to the knowledge mismatching or semantic errors.
   Therefore, granularity tuning on keyword extraction and subsequent semantic calibration deserve further investigation to tackle such randomness.
    
    \textbf{Inactive Sharing of Background Knowledge and Personal Preferences:}
    Since the prerequisite of customized AIGC and SemCom services mainly lies in the proactive sharing of users' personal preferences and background knowledge, devising a scores- or rewards-based incentive mechanism, such as delegated proof of stake-based blockchain, is necessary to attract users to spontaneously contribute personal data to the upgrading of GAI-SCN, where potential reward alternatives include social welfare and tech benefits, etc.
  	
	\section{Conclusions}
	This article explored the potential of applying AIGC into SemCom for service provisioning, where we first showcased the development of SemCom and GAI technologies with their integration cases, and then proposed the GAI-SCN framework.
	Specially, the collaborative cloud-edge-mobile structure was well-devised to incorporate both global and local GAI models with the JSCC process, which not only enables efficient and high-quality meaning delivery, but also significantly reduces transmission traffic as well as latency.
	Moreover, implementation alongside initial simulations was provided, followed by associated open issues and corresponding solutions.
	We hope that our GAI-SCN serves as a pioneer in facilitating communication resource usage as well as user experience for futuristic context-aware and GAI-based wireless SemCom networks.
	
	\section{Acknowledgments}
	This work was supported in part by UK Department for Science, Innovation \& Technology (DSIT) Towards Ubiquitous 3D Open Resilient Network (TUDOR) Project, and in part by the National Research Foundation, Singapore, and Infocomm Media Development Authority under its Future Communications Research \& Development Programme, Defence Science Organisation (DSO) National Laboratories under the AI Singapore Programme (FCP-NTU-RG-2022-010 and FCP-ASTAR-TG-2022-003), Singapore Ministry of Education (MOE) Tier 1 (RG87/22), and the NTU Centre for Computational Technologies in Finance (NTU-CCTF).
	
	\bibliographystyle{IEEEtran}
	\bibliography{main}
	\vspace{-20pt}
	\begin{IEEEbiographynophoto}{Le Xia} (xiale1995@outlook.com)
	obtained his Ph.D. degree in Electronics \& Electrical Engineering from the University of Glasgow, United Kingdom, in 2024. Before that, he received his B.Eng. and M.Eng. in Electronics and Communication Engineering from the University of Electronic Science and Technology of China in 2017 and 2020, respectively. His research interests include next-generation wireless networking, semantic communications, resource optimization, AI for wireless, and smart vehicular networks.
	\end{IEEEbiographynophoto}
	\vspace{-20pt}
	\begin{IEEEbiographynophoto}{Yao Sun} (Yao.Sun@glasgow.ac.uk)
	is currently a Lecturer with the James Watt School of Engineering, the University of Glasgow, UK. His research interests include semantic communications, intelligent wireless networking, and wireless blockchain system.
	\end{IEEEbiographynophoto}
	\vspace{-20pt}
	\begin{IEEEbiographynophoto}{Chengsi Liang} (2357875l@student.gla.ac.uk)
	is currently pursuing her Ph.D. degree with the James Watt School of Engineering, University of Glasgow, UK. Her research interest includes semantic communication and networking.
	\end{IEEEbiographynophoto}
	\vspace{-20pt}
	\begin{IEEEbiographynophoto}{Lei Zhang} (Lei.Zhang@glasgow.ac.uk) 
	is a Professor at the University of Glasgow. He has academia and industry combined research experience on wireless communications and networks, and distributed systems for IoT, blockchain, autonomous systems. He is the founding Chair of IEEE Special Interest Group on Wireless Blockchain Networks in Cognitive Networks Technical Committee.
	\end{IEEEbiographynophoto}
	\vspace{-20pt}
	\begin{IEEEbiographynophoto}{Muhammad Ali Imran} (Muhammad.Imran@glasgow.ac.uk) 
	is a Professor of communication systems with the University of Glasgow, UK, and a Dean with Glasgow College UESTC. He is also an Affiliate Professor with the University of Oklahoma, USA, and a Visiting Professor at University of Surrey, UK. He has over 20 years of combined academic and industry experience with several leading roles in multi-million pounds funded projects.	\end{IEEEbiographynophoto}
	\vspace{-20pt}
	\begin{IEEEbiographynophoto}{Dusit Niyato} (dniyato@ntu.edu.sg)
	is a professor in the School of Computer Science and Engineering, at Nanyang Technological University, Singapore. He received Ph.D. in Electrical and Computer Engineering from the University of Manitoba, Canada in 2008. He has published more than 400 technical papers in the areas of wireless and mobile computing, sustainability, edge intelligence, decentralized machine learning, and incentive mechanism design. He was a Distinguished Lecturer of the IEEE Communications Society from 2016 to 2017. He is a Fellow of IEEE.
	\end{IEEEbiographynophoto}
\end{document}